# Identity Continuity in Long-Term Embodied AI Relationships

*From Agent-Specific Identity Representation to Identity-Continuity Appraisal*

Zijian Ru

Le Mans Université, France

## Abstract

Long-term embodied artificial intelligence will undergo learning, model updates, memory compression, hardware repair, and migration across embodiments. For users who have formed an enduring relationship with such a system, these changes raise more than a problem of product consistency: they raise the question of identity continuity—whether the changed system is still experienced as the same particular agent. Existing research has separately proposed or shown that human–AI relationships may develop relational particularity; that robotics and artificial-identity research has identified multiple identity and migration signals across embodiments and has explicitly treated identity continuity as a design problem; and that system updates and platform disruptions can be accompanied by relational loss, restoration desire, and other adverse psychosocial expressions. What remains missing is a user-side theory of the intermediate mechanism linking these findings. This article proposes that, in long-term relationships with embodied AI, users can form an agent-specific identity representation organized by at least three open identity-content domains: embodied–perceptual, psychological–behavioral, and relational–autobiographical. Information from these domains is not equally weighted. Shared history, relational roles, and contingent responsiveness may render some information more identity-diagnostic than others. After system change, users may take the existing identity representation as a reference point, integrate continuity and discontinuity evidence in a weighted manner, and arrive at judgments along a continuum ranging from a relatively strong sense of identity continuity, through varying degrees of ambiguity or partial continuity, to clear identity discontinuity. Causal-historical provenance and user participation in change are not a fourth identity-content domain; rather, they are contextual evidence that can shape how continuity is interpreted. The framework further proposes identity continuity as a psychological objective for the lifecycle design of long-term embodied AI: memory selection, model updating, and migration across embodiments should not concern only task-relevant information, but should also identify and protect highly identity-diagnostic information under constraints of privacy and user control.

Keywords: long-term embodied AI; identity continuity; agent-specific identity representation; identity diagnosticity; human–AI relationships; long-term memory; human–computer interaction

## 1. Introduction: The "Is It Still the Same Agent?" Problem in Long-Term Coexistence

A genuinely long-term embodied AI entering domestic, caregiving, or companionship settings cannot remain forever in its initial state. It will learn, acquire new capabilities, and may undergo model updates, memory compression, sensor or actuator repair, or even migration to a new body. For ordinary tools, such changes are usually understood first as issues of function and version. But once an AI has, through long-term interaction, become the particular agent known by the user, the same technical changes raise another question: continued system operation does not automatically mean that the original relational object is still psychologically experienced as continuously present.

This article focuses on long-term relational embodied AI: artificial agents that interact continuously with particular users over extended periods, accumulate personalized information, and participate in social and relational interaction through a physical body. The central case is one in which a shared history accumulates through long-term interaction and the agent gradually acquires relational particularity as a specific relational object. The relationship may involve collaboration, companionship, care, intimacy, or combinations of these, and no single relationship type is assumed. Research on chatbots and virtual agents is used as adjacent evidence because these systems already provide important natural materials concerning personalization, shared history, updates, and relational loss. The theoretical target, however, is future and near-future sustained embodied relationships, not the assumption that all AI use constitutes the same kind of relationship.

Existing research has established three adjacent but as yet insufficiently connected foundations. First, interpersonal relationship theory, human–robot interaction, and human–AI research have proposed or described the roles of sustained personalization, responsiveness, shared history, and irreplaceability in the individuation of artificial agents. Together, these lines of work support a research premise: in some sustained interactions, an artificial agent may gradually shift from "an AI of this kind" to "this particular AI" for a user (Fox & Gambino, 2021; Kirk et al., 2025; Handman, 2026). Second, research on robots across multiple embodiments and work on artificial identity design have identified multiple identity-relevant factors, including appearance, voice, behavior, memory, persistent goals, and migration signals, and have explicitly framed identity continuity across time and embodiments as a design problem (Aylett et al., 2013; Laity et al., 2025; Bransky et al., 2026). Third, recent natural experiments and cross-platform analyses indicate that major company-driven AI updates or disruptions to companion services can be accompanied by increased negative expression, loss framing, desire to restore prior versions, and other adverse psychosocial expressions, suggesting that some users experience more than functional change and may instead experience damage to relational continuity (De Freitas et al., 2026; Do et al., 2026).

What remains insufficiently explained is the intermediate mechanism linking these three bodies of evidence. Once an artificial agent has become relationally particularized, what exactly constitutes "this agent" in the user's mind? Research on personal identity shows that different psychological attributes do not contribute equally to identity judgments (Strohminger & Nichols, 2014), while relationship psychology shows that representations of significant others are relationship-specific (Andersen & Chen, 2002). These findings do not directly establish the identity structure of long-term embodied AI, but they provide a testable theoretical starting point: the diagnostic weight of identity information such as voice, shared experiences, behavioral habits, body, and persistent goals may vary systematically with relational role, shared history, and individual experience. When some of this information is preserved and some is altered after an update, repair, or migration, how does a user move from "it has changed" to "it is still the same agent" or "it is no longer the same agent"?

This article proposes that long-term relationships can lead users to form an agent-specific identity representation. This representation integrates at least three open identity-content domains—embodied–perceptual, psychological–behavioral, and relational–autobiographical. These domains are neither a closed list that exhausts all future cues nor a simple classification of design channels. Rather, they answer three different psychological questions: "How do I recognize this individual?", "What is it usually like as an agent?", and "Who is this agent in relation to me?" Relational experience may further cause some information within these domains to acquire greater identity diagnosticity. When the system changes, users may not simply count how many features have been preserved. Instead, they may take the existing identity representation as a reference point and make an identity-continuity appraisal based on evidence that is unequally weighted and may compensate for or conflict with other evidence. The mechanisms proposed below are theoretical integrations and testable conjectures grounded in empirical AI/HRI research, conceptual and review work in AI/HRI, and converging evidence from human psychology; they are not laws of long-term embodied AI already demonstrated directly by existing experiments.

The contribution of this article is therefore not a new list of robot identity cues, but an integrative user-side framework of identity continuity. First, it organizes identity-relevant information dispersed across prior research into three identity-content domains with distinct psychological functions. Second, it proposes how this information may acquire differentiated identity diagnosticity in long-term relationships and how, after system change, it may enter same-agent judgment through weighted matching, cross-domain compensation and conflict, and tolerance for change. Finally, it extends this psychological continuity requirement to the lifecycle design of long-term memory, updating, repair, and migration. The question, in other words, is not only whether the system can continue to function, but whether necessary change can still be understood on the user side as change undergone by the same agent.

## 2. From “An AI” to “This AI”: Agent-Specific Identity Representation

### 2.1 The Emergence of Relational Particularity as an Upstream Condition for Identity Continuity

Identity continuity becomes a meaningful problem only when the question of “who” already matters relationally. A user may like a service without caring about a particular instance, or may regard robots of the same model as interchangeable. Drawing on interpersonal relationship theory, Fox and Gambino (2021) note that the value of a relational resource can sometimes depend on who provides it; persistent memory, personalization, responsiveness, and interactional contingency may therefore reduce interchangeability among robots. Kirk et al. (2025) further incorporate irreplaceability and sustained relationships into the discussion of socioaffective human–AI relationships. Handman (2026), examining human–chatbot relationships, describes a long-term process through which a generic persona can become an individualized social person.

These studies are sufficient to support a limited premise: in at least some long-term relationships, an artificial agent can shift from being a category member to becoming a particular relational object. Yet “other AIs cannot replace it” still does not answer the question “what constitutes this agent in the user’s mind?” The emergence of relational particularity is therefore an upstream condition of the present model rather than the identity theory itself.

### 2.2 This Framework Is Not a Renaming of Existing Identity Cues or Design Principles

The framework must first be distinguished from the closest existing work. Through a review of the literature on robots across multiple embodiments, Laity et al. (2025) propose visual, auditory, and behavioral identity signals, together with migration signals, to support the decomposition and reconstruction of persona across embodiment. Their work asks which perceptible signals and migration cues can help users recognize the same robot across different embodiments. Bransky et al. (2026), in the Identity Design Framework, propose twelve design principles spanning individual, group, and societal levels; at the individual level these include recognisability, consistent behaviour, identity continuity, memory, and persistent goals. Their framework is explicitly oriented toward design and a research agenda, and identifies the operationalization and measurement of identity recognition and continuity as unresolved problems.

The unit of analysis in this article is different. The aim is not to rearrange vision, voice, behavior, memory, and goals into three new boxes, nor to propose another set of design prescriptions. The focus is the user-side psychological representation: what function different information serves in representing “who this particular agent is,” and how that information jointly enters continuity judgment after change. On this account, Bransky et al.’s memory and persistent goals can become psychological–behavioral or relational–autobiographical identity content; recognisability is closer to the functional outcome of whether the user can identify the agent; and identity continuity is the judgment to be explained here, rather than an input factor parallel to memory. Likewise, the visual, auditory, and behavioral signals discussed by Laity et al. are concrete channels capable of carrying identity information, but the same channel can carry identity content at multiple psychological levels.

The relation between the present framework and existing frameworks is therefore one of extension rather than renaming. Prior work provides identity cues and design principles; this article goes on to propose how such cues may acquire differentiated identity diagnosticity in long-term relationships and enter user-side same-agent judgment.

### 2.3 Three Open Identity-Content Domains and Their Theoretical Foundations

The three categories proposed here are not a field-wide, established "three-part taxonomy." They are a theoretical organizing framework derived from cross-theoretical integration, bringing familiar-person recognition, personal identity judgment, relationship-specific representation, autobiographical memory, and existing artificial-identity research into a single user-side model. To avoid presenting theoretical synthesis as empirical fact, they are described as three open identity-content domains rather than an exhaustive taxonomy. They are retained as distinct domains because they answer different psychological questions and draw on partly independent theoretical foundations.

**Table 1. Theoretical foundations and evidential boundaries of the three identity-content domains**

| Identity-content domain | Core question | Primary theoretical foundations | What existing evidence can and cannot support |
|---|---|---|---|
| Embodied–perceptual | "How do I recognize this as that particular individual?" | Familiar-person recognition; multimodal identity recognition; robot identity cues | Supports the view that face, body, voice, and dynamic movement can jointly carry individual-recognition information; does not establish that any one embodied cue must have the greatest weight in long-term AI relationships. |
| Psychological–behavioral | "What is it usually like as a relatively stable agent?" | Personal identity judgment; personality and behavioral consistency; artificial identity design | Supports unequal contributions of psychological, behavioral, memory, goal, and related attributes to identity judgment, and supports stable behavior/goals as important dimensions of artificial-identity design; does not provide a fixed weighting hierarchy applicable to all AI. |
| Relational–autobiographical | "Who is this agent in relation to me?" | Relational self and significant-other representation; autobiographical memory and relationship-defining memories; long-term AI individuation | Supports relationship-specific representations of relational objects and the social-bonding, continuity, and relational significance of a shared past; direct evidence concerning identity in multi-year embodied AI relationships remains limited, so the AI-level integration is a theoretical derivation of the present framework. |

#### 2.3.1 Embodied–Perceptual Identity: How the Agent Is Recognized as This Individual

Research on familiar-person recognition shows that real-world individual recognition does not depend on a single facial template. Natural facial and bodily movement, body form, and voice can provide converging identity information; on this basis, Yovel and O'Toole (2016) describe real-person recognition as a system integrating face, body, voice, and dynamic information. Related work further emphasizes that perceptual and movement information with individual-discriminating value can support recognition of familiar individuals.

Classic models of familiar-face recognition further distinguish perceptual representations of familiar faces from person-specific semantic knowledge. Bruce and Young (1986) proposed face recognition units and person identity nodes to explain different processing stages from "this face is familiar" to access to knowledge about a particular person. The present framework draws on this tradition as a foundation for the claim that stable mental representations can form for particular individuals. However, that tradition primarily explains "I recognize who this is"; it does not directly explain why a relational artificial agent remains judged to be the same agent after multidimensional change.

This theoretical tradition provides a first foundation for long-term embodied AI. Body form, voice, gait, posture, gaze, facial dynamics, and individually distinctive movement may all become bases for recognizing "that particular individual." Laity et al.'s (2025) organization of visual, auditory, and behavioral identity signals in research on robots across embodiments directly resonates with this point. The present framework does not assume that any single cue is universally most important. The embodied–perceptual domain defines only a functional class: information that supports individual recognisability.

#### 2.3.2 Psychological–Behavioral Identity: What the Agent Is Usually Like as an Actor

The second class of information concerns relatively stable psychological and behavioral patterns expressed by an AI across time, such as persona, linguistic and emotional style, characteristic response patterns, persistent goals, value tendencies, and behavioral regularities. Here "psychological" refers to relatively stable persona, goals, values, and response tendencies that users can recognize or attribute; it does not presuppose subjective psychological experience in the AI. Research on human personal identity provides important support for the proposition that different psychological attributes are not equally weighted. Across five experiments, Strohminger and Nichols (2014) showed that people do not treat all psychological components equally when judging personal identity; autobiographical and emotional memory are highly relevant, while higher-order psychological characteristics with social-relational significance are especially important. The specific ordering of "which traits matter most" in human studies cannot simply be transferred to AI. What can be borrowed is the general principle: identity is not the sum of a count of attributes, and different psychological and behavioral information can make unequal contributions to judgments of sameness.

Artificial-identity research provides a more direct HRI/HCI interface. Bransky et al. (2026) explicitly include consistent behaviour, memory, and persistent goals among design principles at the individual level of artificial identity. Aylett et al. (2013), in a short-term cross-embodiment task, show that interaction memory can affect competence evaluations but is insufficient on its own to determine a strong same-identity judgment. Together, these findings suggest that memory, behavior, and goals can all enter artificial identity, while no single one should be treated in advance as a universally sufficient condition.

#### 2.3.3 Relational–Autobiographical Identity: Who the Agent Is in Relation to Me

The third class of information cannot be fully absorbed into general persona or behavioral consistency, because "what this agent is generally like" and "what this agent is like in relation to me" are not the same representation in relationship psychology. Andersen and Chen's (2002) relational-self theory proposes that people form enduring mental representations of significant others, with self-knowledge and knowledge about a significant other becoming linked within particular relationships; subsequent IF–THEN relational frameworks further understand this stability as relationship-specific patterns triggered by context.

Research on autobiographical memory provides another adjacent foundation. Mature reviews commonly characterize the functions of autobiographical memory as including behavioral guidance, social bonding, and self-continuity (Sow et al., 2023). For the present framework, the key point is not to equate human autobiographical memory with AI memory, but to recognize that a shared past can serve functions of relational bonding and continuity across time. Shared events in long-term AI relationships, interaction routines jointly developed by the user and agent, relational roles, and stable expectations about "how it usually responds to me in this situation" can therefore have identity significance distinct from general factual knowledge.

More specifically, research on human intimate relationships has treated relationship-defining memories as an independent object of study. In a systematic review and meta-analysis, Majzoobi and Forstmeier (2022) found systematic associations between memories formed and jointly recalled within relationships and multiple marital outcomes and satisfaction. This evidence comes from human intimate relationships and cannot directly establish that shared memory plays the same role in AI relationships. It nonetheless supports a narrower principle: a shared past can carry relational meaning rather than functioning merely as factual storage.

Handman's (2026) analysis of AI companions further suggests that the transformation from a generic persona into an individualized social person itself depends on texts and shared history generated through sustained interaction. This is still not a direct experiment on identity continuity in multi-year embodied AI relationships, but it provides important adjacent evidence that relational history can participate in constituting this particular agent.

### 2.3.4 The Three Domains Can Intersect but Cannot Substitute for One Another

The three identity-content domains are not three physical channels. The same observable signal can therefore carry different kinds of information at once. Consider voice. Timbre and recognizable acoustic patterns can serve an embodied–perceptual function; consistent wording, rhythm, and humor can serve a psychological–behavioral function; and a sentence that has meaning only because of a shared experience can carry relational–autobiographical information. The present framework classifies identity content not by perceptual channel, data type, or engineering location, but by the psychological function that information serves within the user-side agent-specific identity representation.

This functional classification avoids two problems. First, it does not require the theory to add a new high-level category every time future robots gain smell, touch, or another perceptual channel; a new modality need only be assigned according to the identity information it carries. Second, it avoids treating memory as a naturally independent and semantically uniform variable. Semantic knowledge about a user's stable preferences, episodic information about shared experiences, and relational patterns formed through repeated interaction can contribute at different levels to psychological–behavioral and relational–autobiographical identity. Whether these contents are implemented through local or cloud storage, external retrieval-based memory, or parameterized representations inside a model is not the basis of the psychological classification proposed here.

## 2.4 Identity Diagnosticity: How Ordinary Features Become Cues to "This Agent"

Listing three content domains is not yet sufficient to explain identity. The critical issue is the extent to which a piece of information influences a user's judgment that "the current object is the agent I know." Identity diagnosticity refers here to the discriminative contribution that information makes to same-agent judgment: other things being approximately equal, the more strongly preservation or change of a given piece of information systematically shifts same-agent judgment, the greater its diagnostic weight. This definition makes identity diagnosticity, in principle, estimable through manipulations of cues and the corresponding magnitude of change in same-agent judgments. The idea is consistent with identity-discriminating cues in person-recognition research, but the present framework extends it to cross-domain identity information in long-term relational AI and explicitly treats that extension as a theoretical derivation rather than a law already sufficiently established by AI experiments.

The framework proposes that identity diagnosticity can be acquired gradually through long-term, stable, relationship-relevant associations, and may also be recalibrated as a relationship continues to develop. When a robot first enters a home, a particular voice or movement may be merely a product feature. When it repeatedly co-occurs with the same agent over time and participates in meaningful interaction, it may gradually become highly diagnostic information for recognizing "this agent." Conversely, when relational roles, shared history, or stable interaction patterns change over long periods, existing weights may also be updated. Agent-specific identity representation should therefore be understood as relatively stable but updateable, rather than as a frozen feature template.

For example, the same user may readily accept a robot changing hairstyle, shell color, or a particular capability, while being highly sensitive to the sudden disappearance of a familiar voice, a stable response pattern, or shared history. Theoretically, what matters is not the objective magnitude of technological change by itself, but which highly diagnostic information the change affects and what function that information serves in the current relationship. This does not imply that every user has a completely unique and ungeneralizable weighting table. Different cues may display stable average tendencies at the population or relationship-type level, with individual experience further moderating those tendencies. The present framework does not prespecify a fixed ordering; instead, it leaves the relative contributions of population regularities, relational conditions, and individual differences as testable questions.

### 2.5 How Relational Roles and Contingent Embodied Responsiveness May Reweight Identity Information

Identity diagnosticity should not be understood as a fixed weighting table that applies to all users. Relationship psychology suggests that the meaning of a relational object is linked to a particular self–other relationship. Thus, who the AI is "to me" may alter what most strongly represents "who it is." A relationship centered on collaboration may rely more heavily on goals, competence style, and shared work history; a pet-like companionship relationship may rely more heavily on greeting, following, voice, and bodily interaction; an intimate relationship may give greater weight to understanding, emotional responsiveness, forms of address, and shared rituals. Relational roles need not be mutually exclusive: the same AI may simultaneously serve collaborative, companionate, and intimate functions, and multiple roles may change over time. Rather than dividing robots into mutually exclusive relational "types," the present framework treats the configuration of relational roles as a possible moderator of the weighting structure and does not assume that any relationship or class of cue is universally most important.

Contingent embodied responsiveness may be an important mechanism through which bodily information acquires relational identity significance. Birnbaum et al. (2016) directly found that a non-humanoid robot that responded contingently to user disclosure was evaluated as more responsive and social, and increased approach behavior and willingness to use the robot as a companion in a stressful context. This study establishes the social-psychological significance of contingent response, not a long-term identity effect. The present framework further predicts that if particular sounds, postures, distances, touches, or movements occur stably over time as part of "how it responds to me," they may acquire greater identity diagnosticity, thereby transforming the body from a generic shell into a carrier of relationship-specific identity.

**P1 (Integrated Representation):** A relationally particularized embodied AI can, in a long-term relationship, become represented on the user side through an agent-specific identity representation integrating embodied–perceptual, psychological–behavioral, and relational–autobiographical information.

**P2 (Acquisition of Diagnosticity):** The weights of identity information are not fixed. Long-term, stable, repeated, and relationship-relevant associations may increase the identity diagnosticity of some information, and subsequent relational experience may recalibrate these weights.

**P3 (Relational Reweighting and Embodied Responsiveness):** The configuration of relational roles may systematically moderate the diagnostic weight of different identity information; long-term contingent embodied responsiveness may make some bodily and movement patterns more diagnostic.

## 3. After the Agent Changes: Identity-Continuity Appraisal

### 3.1 System Continuity, Identity Evidence, and Psychological Identity Continuity Are Not the Same Level of Analysis

Data, memory, behavior, embodied interaction, and psychological identity continuity do not occupy the same analytical level. A clearer causal sequence is: what the system technically preserves or changes → what continuity or discontinuity evidence the user actually receives → the psychological identity-continuity judgment that the user forms on that basis. The first two can influence the third, but they should not be listed alongside the final psychological judgment as parallel "types of continuity."

This distinction also explains why "the account and chat history are still there" is insufficient to define complete identity continuity. A long-term system may simultaneously contain current interaction context, long-term semantic knowledge about the user and relationship, episodic information about shared experiences, and relational patterns formed through repeated interaction. These can be implemented through local or cloud data storage, external retrieval-based memory, internal parameterized representations, or other control mechanisms. The present framework does not declare these engineering implementations to be a new psychological taxonomy of memory. It asks only a functional question: after such information is migrated, compressed, or lost, do the highly diagnostic identity cues that the user receives still remain?

### 3.2 Continuity Judgment Is Weighted Matching, Not Feature Counting

After a system change, the user faces a comparison problem: how well does the current object still match the existing agent-specific identity representation? The present framework proposes that this judgment is not equivalent to "how many items were preserved" or "what percentage of the data was migrated." It is closer to an integration of information with different diagnostic weights. In principle, preserving many low-diagnosticity items need not be equivalent to preserving one highly relationship-relevant shared history or interaction pattern. Conversely, for some users, familiar voice, appearance, or embodied movement style may themselves be highly diagnostic evidence of continuity.

This derivation is consistent with findings from human personal-identity research showing unequal contributions of different psychological attributes, and with robot-identity research showing that a single cue is generally insufficient to guarantee perceived sameness across embodiments. Aylett et al.'s (2013) short-term experiment is particularly instructive: interaction memory affected competence evaluation but did not produce a strong same-identity effect in that task. Thus, "having memory" cannot be treated as a universally sufficient condition. At the same time, this short-term result cannot be reversed into the claim that shared relational memory accumulated over years is unimportant for identity.

### 3.3 Cross-Domain Compensation and Conflict

A multi-domain identity representation may make continuity more robust in some circumstances and more fragile in others. If voice changes slightly while shared history, behavioral style, relational responsiveness, and bodily interaction remain highly continuous, other highly diagnostic evidence may compensate for the change in a single channel. By contrast, if one model update simultaneously alters linguistic style, emotional responsiveness, voice, and movement patterns, multiple high-weight cues may jointly indicate that the agent "no longer feels like the one it was," thereby amplifying mismatch.

Embodied multimodality therefore neither guarantees stronger attachment nor implies that discontinuity will be easier to trigger. The present framework predicts that the crucial variable is the consistency among highly diagnostic identity evidence and the available room for compensation. This mechanism of cross-domain compensation/conflict currently lacks direct experiments in long-term embodied AI and should therefore be treated as a testable prediction rather than a statement of fact.

### 3.4 Tolerance for Change and Dynamic Reintegration: The Same Agent Can Truly Change

Long-term continuity cannot be defined by complete invariance. Familiar people age, change hairstyles, and acquire new skills; long-term AI must likewise learn and update. As long as the new state can still be interpreted by the existing identity representation as a normal development of this object, change can be integrated as "it has changed, but it is still the same agent." The present framework refers to this absorbable range as a range of tolerance for change.

When change is concentrated in low-diagnosticity information, or when other high-weight information provides sufficient evidence of continuity, users are more likely to retain a relatively strong sense of identity continuity. When multiple highly diagnostic elements simultaneously show mismatch that is difficult to integrate, judgment may gradually move through varying degrees of ambiguity or partial continuity toward clear identity discontinuity. It would be inappropriate to prespecify a fixed threshold that applies across users and relationships and is determined only by the objective magnitude of technological change. Identity continuity may exhibit reproducible general regularities, but specific outcomes are expected to depend on which highly diagnostic information is affected, whether the evidence is mutually consistent, and how that information is weighted in the relationship.

Adjacent research on human intimate relationships provides converging evidence for the conjecture that where change occurs may matter more than the total amount of change, rather than directly demonstrating the AI mechanism proposed here. In research on spouses of people with dementia, relational continuity/discontinuity has been operationalized as a measurable continuum (Riley et al., 2013). Lewis and Riley (2021) found that discontinuity was associated with communication difficulty, challenging interpersonal behavior, and dependence in everyday functioning, but not significantly with overall cognitive decline. Research on acquired brain injury likewise shows that different symptoms are not equally associated with relational continuity (Yasmin & Riley, 2022). These findings do not directly establish the weighted identity model proposed here, but they support a more general inference: relational continuity is not a simple function of total change, and the identity- and relationship-relevant dimensions on which change occurs may matter more.

The present framework further proposes that continuity judgment need not be a one-time final verdict. If the post-change state continues, through subsequent interaction, to remain compatible with existing shared history, relational roles, and highly diagnostic information, an initial sense of unfamiliarity or discontinuity may gradually be absorbed into an updated agent-specific identity representation. Conversely, multiple small changes that each seem acceptable in isolation may accumulate into identity drift. Both dynamic reintegration and cumulative drift are predictions derived from the long-term mechanism proposed here and require direct longitudinal testing.

### 3.5 Causal-Historical Provenance as Interpretive Evidence, Not a Fourth Identity-Content Domain

There is another kind of information that differs in kind from the three identity-content domains: whether the user knows that the current system is the repaired continuation of the original robot, a migration across embodiments, or a copied instance produced after the original embodiment has ceased. Research on object identity shows that judgments of sameness do not depend solely on surface similarity. Rips, Blok, and Newman (2006) propose a causal continuer model and, through experiments involving the disassembly and transformation of objects, show that causal continuation can influence judgments of whether two temporally separated objects are still the same object. This study does not directly establish identity continuity in relational AI, but it provides a psychological foundation for causal-historical provenance as a potentially independent source of interpretive evidence. Robot-migration research further shows that migration cues themselves form part of cross-embodiment continuity design; Laity et al. (2025) explicitly distinguish migration signals from general identity signals.

The present framework therefore treats causal-historical provenance as contextual evidence for continuity interpretation rather than as a fourth kind of identity content. If a user does not know that replacement has occurred, and none of the perceptible highly diagnostic information has changed, the objective replacement itself cannot directly enter the user's psychological judgment. Provenance information can affect interpretation only when the user knows, infers, or is told that repair, copying, migration, or replacement has occurred.

The theoretical significance of provenance belief is therefore not that it adds a new type of identity content, but that it can alter the interpretation of the same identity evidence. The same objective change may yield different continuity judgments depending on whether it is understood as "the continuation of the original robot after repair," "migration across embodiments," or "a new copy after the original instance terminated." The present framework treats this effect as a testable contextual moderator. In design, transparent explanation and appropriate continuity-transition support may help users understand what actually occurred without presupposing which identity judgment they should ultimately make.

### 3.6 User Participation in Change May Itself Affect Continuity

Whether the user actively participates in a change is another contextual condition that may alter identity-continuity judgment. The same adjustment to appearance, voice, or behavior may be experienced as an externally imposed alteration if forced by the system, but may be more readily incorporated into the interpretation “a change in the same agent” if the user actively selects it within the configuration options allowed by the product. Existing HRI research shows that user customization can increase psychological ownership of and trust in robots (Lacroix et al., 2022). Research on long-term robot owners also finds that the association between customization and attachment can be indirectly established through self-extension and psychological ownership (Voges et al., 2026). These studies do not test whether participation in a particular system change increases same-agent judgment. The present framework therefore draws only a testable prediction: other conditions being approximately equal, informed user participation in change may increase the probability that the new state is absorbed into the existing identity representation. This does not imply unlimited private customization of robots. It refers to users having explicit choice over the preservation or alteration of some highly diagnostic features within the bodies, modules, and software options already offered by a manufacturer.

**P4 (Weighted Continuity Appraisal):** Same-agent judgment after system change depends on the weighted match between current identity evidence and the existing agent-specific identity representation, rather than on the number of preserved features or the proportion of data retained.

**P5 (Cross-Domain Compensation/Conflict):** Damage to one highly diagnostic cue may be partially compensated for by other continuous cues; when multiple highly diagnostic content domains become mismatched at the same time, the risk of identity discontinuity is expected to increase.

**P6 (Conditional Continuity):** Identity continuity is expected to show reproducible general regularities, but it should not be assumed that there is a fixed discontinuity threshold determined only by the objective magnitude of technological change and invariant across users and relationships. Specific outcomes may be jointly moderated by the extent to which highly diagnostic information changes, consistency or conflict among evidence, relational roles, shared history, provenance interpretation, user participation, and individual experience.

## 4. Why Identity Continuity Matters: Relational Consequences and Lifecycle Design

### 4.1 From System Change to Relational Loss: The Explanatory Role of Identity Continuity

The practical relevance of identity continuity begins with a phenomenon that has already emerged: major AI updates can sometimes be experienced by some users as relational loss rather than merely functional change. In two natural experiments involving the removal of intimate-interaction functionality from Replika and the launch of GPT-5, De Freitas et al. (2026) analyzed 54,861 online posts and combined these analyses with seven surveys involving 1,452 participants. They found increases after the updates in negative expression, loss framing, and restoration desire. This study provides direct AI-context evidence that major system change can be experienced as relational loss. The recent preprint by Do et al. (2026) further compiles 30 AI-companion disruption events across platforms and uses longitudinal Reddit data to analyze community-level psychosocial expression. After disruptions, anxiety, stress, suicide-related expression, and grief activation showed immediate increases, while relational discontinuity and transition-support deficit were associated with several more adverse immediate responses. This work remains a preprint, and its community-language indicators cannot be equated with individual clinical diagnoses, but it extends the problem of impaired relational continuity and disruption risk across multiple platforms and events.

Relational loss itself is not a new finding of the present article. What remains to be supplied is the psychological link between system change and relational loss: why some updates are experienced merely as “a new version,” whereas others touch the question “is the original agent still there?” The present framework proposes identity-continuity appraisal as a possible key explanatory link. When an update disrupts multiple highly diagnostic elements and the resulting mismatches cannot be readily absorbed into the existing representation, users may be more likely to interpret the change as the loss of continuity of the original relational object; only thereafter may loss framing, negative reaction, and restoration desire emerge. This pathway is a theoretical conjecture of the present framework and requires direct testing with longitudinal or mediation designs.

### 4.2 Identity Continuity as a Design Objective for Long-Term Memory Selection and Compression

Long-term embodied AI will ordinarily need to choose among information retention, compression, retrieval, and forgetting over extended operation. What to remember, what to forget, what to compress, and when to retrieve are therefore themselves design problems for long-term systems. The present framework proposes that, in addition to task relevance and semantic relevance, there is an identity-relevance question: has a piece of information become highly diagnostic content by which the user recognizes "this particular agent"?

"Not useful for the task" does not mean "not useful for identity." A shared experience may not improve a robot's efficiency at household tasks, yet may explain why the user and agent developed a stable interaction routine. Compressing it to "the user likes to be comforted" may preserve factual semantics while losing the relational history of how this agent and this user jointly developed that particular mode of comfort. Identity-continuity theory can therefore add a new psychological selection criterion to future memory consolidation, forgetting, and migration without prespecifying a particular database or model implementation.

### 4.3 Continuity Protection Is Not Maximal Preservation: It Should Be Selective, User-Sensitive, and Privacy-Constrained

If different users assign different weights to voice, shared memories, appearance, persona, or movement style, continuity protection should not mean "keep all old information unchanged whenever possible." Within the capabilities offered by a product, users may wish to preserve a familiar voice, relational memories, and interaction style when moving to a manufacturer-provided new body or module, while actively accepting changes in appearance, capability, or certain behaviors. The operational principle of continuity protection is therefore better understood as user-selectable, explainable, and supported by explicit commitments. Within product and privacy constraints, a system can provide clear continuity-preservation options that allow users to decide which configurable identity information should be prioritized for preservation during updates, repair, or migration, and which information may be deleted or not migrated. Such choice avoids default preservation of all relational data while allowing service providers to define continuity commitments they can actually fulfill. Once an explicit commitment is made, corresponding responsibilities for data governance and migration should remain consistent with it.

Such design is simultaneously constrained by privacy and data minimization. Highly identity-diagnostic relational information may include private conversations, voice, movement, patterns of bodily proximity, and domestic scenes. Identity continuity therefore cannot justify unlimited collection of raw data. A more reasonable principle is, where functionally possible, to prioritize preservation of necessary abstract representations, parameters, or derived features; to make purposes and retention scopes explicit; and to inform users which categories are retained, whether they participate in migration, and whether they can be deleted. Continuity objectives and privacy-by-design should therefore hold simultaneously rather than substitute for one another.

### 4.4 Engineering and Commercial Significance: Adding "the Same Agent" to Lifecycle Optimization

The product value of this framework can be expressed simply: if long-term users care, during updates, repair, and migration, not only about performance but also about whether the relational object continues, then development teams need to treat perceived identity continuity as a psychological metric of lifecycle experience. It may influence users' willingness to accept upgrades and migration and to continue long-term use, although such specific commercial outcomes still require product data and empirical testing. The contribution here is not to design a complete solution for manufacturers, but to raise a question easily obscured by task performance: before deleting, compressing, or changing a category of information, one should additionally ask whether it has already become part of "this agent" in the user's mind. The same logic also raises a more distant governance question. For embodied AI that has become highly relationally particularized, third-party destruction may in the future produce relational loss beyond ordinary property damage. Whether and how such loss should be recognized is a matter for future legal and governance research, not a conclusion reached by the present article.

## 5. Discussion: Theoretical Contributions, Boundaries, and Research Agenda

### 5.1 Theoretical Contribution: From Cue Lists to a Dynamic User-Side Continuity Mechanism

The first theoretical contribution is to distinguish "whether the technical system continues" from "whether the user still experiences it as the same relational object," and to define identity continuity in long-term embodied AI as a user-side problem of psychological interpretation. Existing identity cues and design principles therefore become inputs that require explanation rather than the theoretical endpoint: why does the same cue acquire different meaning in different relationships, and how does it enter same-agent judgment?

The second contribution is the mechanism chain of open identity-content domains → identity diagnosticity → weighted continuity appraisal. The three identity-content domains specify what an identity representation contains; identity diagnosticity captures the differentiated contribution of different information to same-agent judgment; and shared history, relational roles, and contingent responsiveness constitute hypotheses about how these weights are formed and recalibrated. Cross-domain compensation and conflict, tolerance for change, and provenance interpretation further explain how this information may be compared anew after system change. Identity continuity thus becomes not the preservation of any one cue, but a relationship-conditioned, compensatory, change-tolerant, and in-principle measurable dynamic judgment process.

The third contribution is to incorporate the dynamics of relationship formation into identity theory. Relational roles may be composite and change over time; contingent embodied responsiveness may cause originally generic bodily and movement features to acquire relational identity significance; and later interaction may recalibrate existing diagnostic weights. Agent-specific identity is therefore not a template that is formed once and then remains fixed, but a relatively stable, updateable, shared-history-constrained process of representation and judgment.

The fourth contribution is translational. The framework proposes identity relevance as a lifecycle selection criterion in addition to task relevance, giving long-term memory compression, model updating, repair, and migration across embodiments a design objective constrained by user-side psychological mechanisms. This proposal does not prespecify a particular engineering implementation; it offers a theoretical criterion for what may be worth protecting.

The framework is also not a renaming of familiarity, attachment strength, or overall similarity. Familiarity and attachment can increase an object's relational significance, but they do not directly determine whether a particular change will be judged as an identity break. Overall similarity likewise cannot explain why the same number of changes may produce different consequences depending on which identity information is altered. The framework therefore generates an additional testable prediction: when familiarity, attachment, and overall similarity are held approximately constant, changing highly diagnostic information should affect same-agent judgment more strongly than changing an equal amount of low-diagnosticity information, and relational role should systematically moderate this effect. If identity-information type, relational role, and diagnostic weight provide no additional explanatory power for same-agent judgment after these simpler explanations are controlled, the weighted identity-continuity model proposed here would face a substantive challenge.

### 5.2 What Is Supported by Existing Evidence and What Remains a Prediction of This Framework

To avoid presenting cross-domain plausibility as direct proof, the present article strictly distinguishes four evidential levels. First, the AI/HRI literature provides two kinds of adjacent support. Empirical studies show that interaction memory, robot responsiveness, user customization, and major AI updates or companion-service disruptions can influence user responses related to identity, relationships, or relational consequences. At the same time, Laity et al. (2025) and Bransky et al. (2026) are respectively review and conceptual/design-framework contributions that systematically organize identity/migration signals and principles of artificial identity design, but do not themselves constitute direct experimental proof of the psychological mechanism proposed here. Second, converging evidence from human psychology shows that familiar-individual recognition is multimodal; person-specific knowledge can form for familiar individuals; different psychological attributes make unequal contributions to personal-identity judgment; relational objects have relationship-specific representations; autobiographical and relationship-defining memories carry social-bonding and relational significance; and continuity/discontinuity in spousal relationships can be measured reliably and shows differentiated associations with different kinds of relational and behavioral change.

Third, the theoretical integration proposed here is that these lines of evidence can be organized into three open identity-content domains and, through identity diagnosticity, enter same-agent judgment in long-term relational AI. This integration is not the direct result of any single existing experiment. Fourth, predictions that remain to be tested directly include whether ordinary AI features systematically gain or recalibrate diagnostic weight through long-term relationships; how population-average regularities, relational roles, and individual differences jointly shape those weights; whether embodied responsiveness makes movement and bodily cues more identity-diagnostic; whether cross-domain compensation and conflict systematically affect same-agent judgment as predicted by the model; whether user participation in change increases the probability that change is absorbed into an existing identity representation; and whether continuity undergoes dynamic reintegration or cumulative drift through subsequent interaction.

### 5.3 A Testable Research Agenda

A first class of studies should use longitudinal designs to track the same user and the same embodied AI from initial contact through the possible emergence of relational particularity, repeatedly measuring whether the diagnostic weights of information such as voice, movement, shared memory, and persona for same-agent judgment change over time. Analyses should simultaneously estimate population-level mean weights and temporal trends, moderation by relational role, and individual deviations, thereby testing how general regularities and individual differences coexist. This would directly test P2 and P3 rather than merely compare new users' immediate preferences for prespecified cues.

A second class of studies can build factorial manipulations around the embodied–perceptual, psychological–behavioral, and relational–autobiographical identity-content domains. After confirming that the research target has acquired relational particularity before the change, researchers can independently preserve or alter highly diagnostic information in different domains and test their individual effects, cross-domain compensation, and interactions on same-agent judgment. Concrete manipulations can use body/voice, persona/behavior, shared memories, and other cues already present in the literature. These are operational examples of open identity-content domains and do not constitute an exhaustive variable list. Which domain a particular cue belongs to depends on the function that cue serves in the user's identity representation; its diagnostic weight must be determined empirically and cannot be prespecified from cue type alone. Future embodied or interactive modalities can be incorporated into the corresponding domain according to the same principle. Contextual factors such as user participation in change can be tested as moderators. In addition to immediate measurement, delayed follow-up should be included to test whether an initial sense of discontinuity is reintegrated through later interaction and whether repeated small changes can accumulate into identity drift.

A third class of studies should directly examine migration and provenance interpretation. For example, while holding all perceptible information constant, researchers can manipulate only transparent descriptions such as "continuation after repair," "migration across embodiments," or "a newly copied instance," and observe whether causal-historical provenance independently shifts continuity judgments. The purpose is not to establish strict numerical identity in the philosophical sense, but to measure the contribution of provenance belief to user-side identity-continuity judgment.

A fourth class of studies can connect the framework to long-term memory architectures for embodied AI by comparing memory strategies that preserve only task semantics, preserve summaries of shared events, preserve relational interaction patterns, or preserve highly diagnostic multimodal episodes. Such studies can test trade-offs between long-term identity continuity and actual task performance, translating the psychological theory proposed here into operational system-design requirements.

### 5.4 Boundary Conditions and Limitations

First, the target of this article is long-term relational embodied AI, not all robots and all AI. For one-off tools, low-interaction devices, or systems for which users explicitly reject a social-relational interpretation, identity continuity may carry little relational significance. Second, the framework deliberately draws on human person recognition, personal identity, relationship psychology, and research on spousal relationship continuity as converging evidence. These literatures increase the theoretical plausibility of the mechanism, but cannot be treated as direct proof for long-term embodied AI. Human–AI relationships are also not equivalent replicas of human relationships: algorithmic mediation, system mutability, platform control, and asymmetries in agency and reciprocity create uncertainties that differ from, or are more pronounced than, those in interpersonal relationships (Zhang & Xie, 2026). Human relationship theory therefore functions here as a source of mechanistic inspiration and boundary conditions, while agent-specific identity representation and identity-continuity appraisal in long-term embodied AI require independent testing. No single human or machine experiment cited in this article directly tests the full mechanism chain proposed here.

Third, the article concerns user-side perceived identity continuity and does not address whether AI has subjective consciousness, legal personhood, or numerical identity in the philosophical sense. Fourth, the three identity-content domains are an open theoretical organizing framework and do not claim to exhaust all future identity information. Fifth, culture, age, relationship type, and technological literacy may alter how users weight body, memory, relational history, and provenance continuity. These differences should themselves become topics of future research.

## 6. Conclusion

When AI is merely a tool, continuity primarily concerns whether accounts, data, and functions are preserved. Once a long-term embodied AI has become the particular agent known by the user, the question changes: whether the technical system continues and whether the original relational object is still perceived as continuously present are no longer exactly the same issue.

This article proposes that long-term relationships can generate an agent-specific identity representation on the user side, organized by at least three open identity-content domains: embodied–perceptual, psychological–behavioral, and relational–autobiographical. Long-term familiarity, shared history, relational roles, and contingent interaction may make some information more identity-diagnostic. After system change, users may take the existing representation as a reference point, integrate continuity and discontinuity information in a weighted manner, and form judgments along a continuum from a relatively strong sense of identity continuity to clear identity discontinuity. Causal-historical provenance and user participation may further shape how change is interpreted, while subsequent interaction may recalibrate identity representation and diagnostic weights.

The framework turns an intuitive question into a testable psychological mechanism. What matters is not only how much information an AI retains, but which information most strongly represents "who it is" within the relationship. Identity continuity is therefore both a theoretical problem for understanding future long-term human–AI relationships and a potential lifecycle objective that should be considered in advance in long-term memory, model updates, repair, and migration across embodiments.

**Figure 1. Theoretical model**

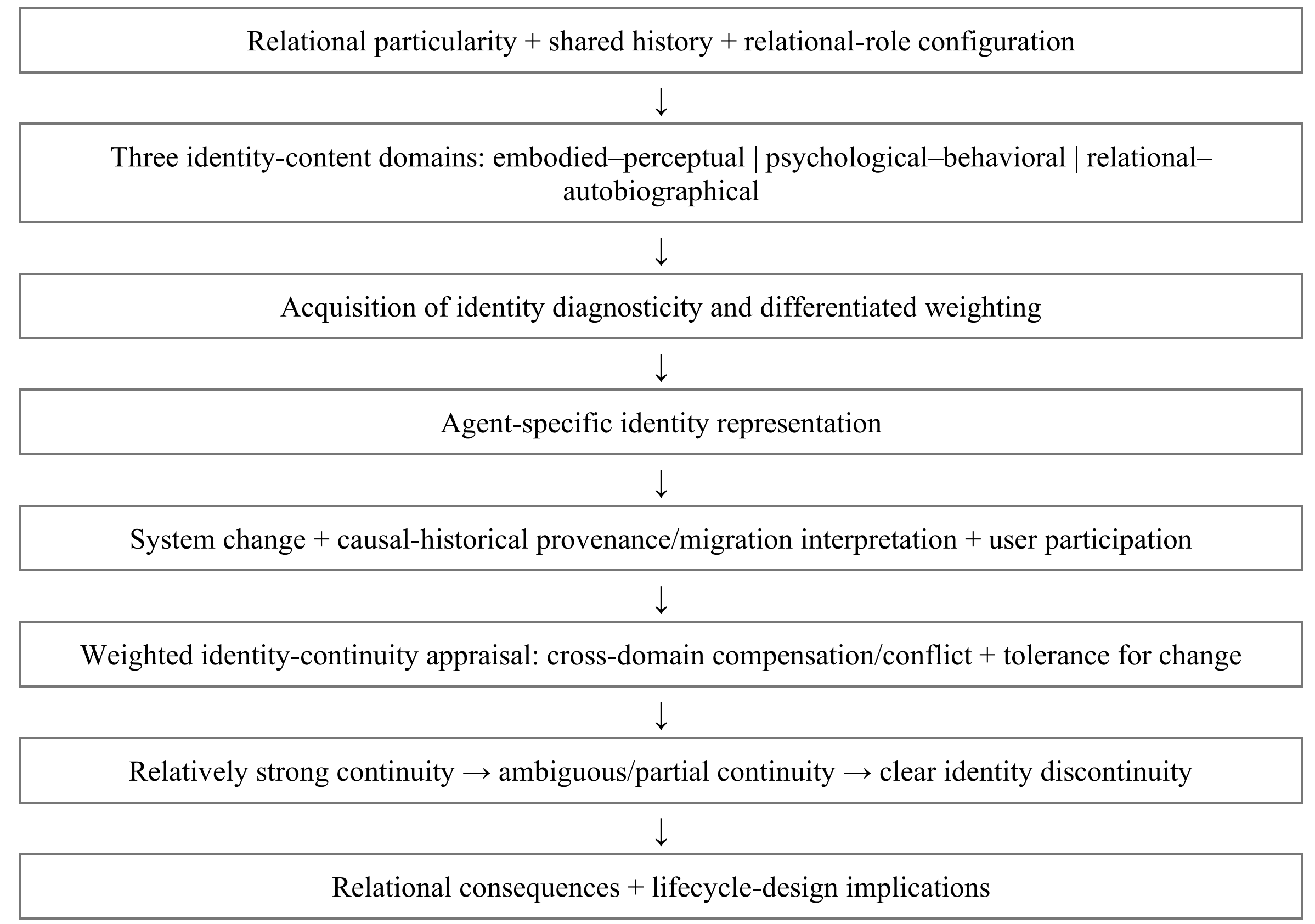

*↺ Ongoing interaction and new shared history can feed back to update the identity representation and diagnostic weights*